\documentclass[a4paper,11pt]{article}
\usepackage{jheppub} 
\usepackage{bm}
\usepackage{lineno}
\usepackage{mathrsfs}

\arxivnumber{2408.16229} 

\title{\boldmath 
Enhanced detectability of  axion's electromagnetic response with a RF-excited magnetic field in cavity}

\author[a]{Li Gao,}
\author[a]{Hao Zheng,}
\author[a]{Xianing Feng,}
\author[a]{Suirong He,}
\affiliation[a]{HgerD Collaboration and Information Quantum Technology Laboratory,
School of Information Science and Technology, Southwest Jiaotong University, Chengdu 610031, China.}
\author[b]{Lingbo Zhao,}
\affiliation[b]{
School of Science, Donghua University,
Shanghai 200051, China}
\author[c]{Qingquan Jiang,}
\affiliation[c]{School of Physics and Astronomy, China West Normal University,
Nanchong 637009, China.}
\author[a]{L. F. Wei}

\emailAdd{ lfwei@swjtu.edu.cn}

\abstract{
Haloscope is one of the typical installations to detect the electromagnetic
responses (EMRs) of axion field in radio-frequency (RF) and microwave bands. Given that the detectable
signals of the usual Haloscope-type detectors (HTDs), biased only by high
stationary magnetic fields, are just the second axion-photon energy and thus are very weak, here we propose a
feasible approach to significantly improve their sensitivity by additionally applying a transverse RF- or microwave modulated magnetic field to excite the cavity's magnetic resonant mode  to produce the first-order axion-photon 
energy response signals. Accordingly, it can be argued that the achievable detection sensitivity
of the upgrading HTD (i.e., UHTD) could be enhanced by {\color{blue}$0.3\sim 1$} orders of magnitude, compared with that achieved by the existing HTDs without the transverse RF-excited magnetic field. The feasibility
of the proposed UHTD is also discussed.}

\begin{document}
\maketitle
\flushbottom


\section{Introduction.}
\label{sec:intro}
Dark matter is believed to be one of the indispensable components of the universe, even though it cannot be directly observed~\cite{NRP2022,A2016}. It is believed that the existence of dark matter could be indirectly tested by probing various effects caused by its possible weak interaction with the particles in Standard Model (SM) at the microscopic level~\cite{ARAA2010,A2024}. However, up to date, no definitive signal of the dark matter has been discovered yet, although a series of observations have provided various relevant parameter constraints. Given the existence of WIMPs (Weakly Interacting Massive Particles) with larger masses has been ruled out almost completely~\cite{PRL2018}, detections of the lighter dark matter candidates, such as dark photons, axions, and axion-like particles, etc. have received increasing attention in recent years~\cite{SA2022}.

\begin{figure}[b]
\centering
\includegraphics[width=0.8\textwidth]{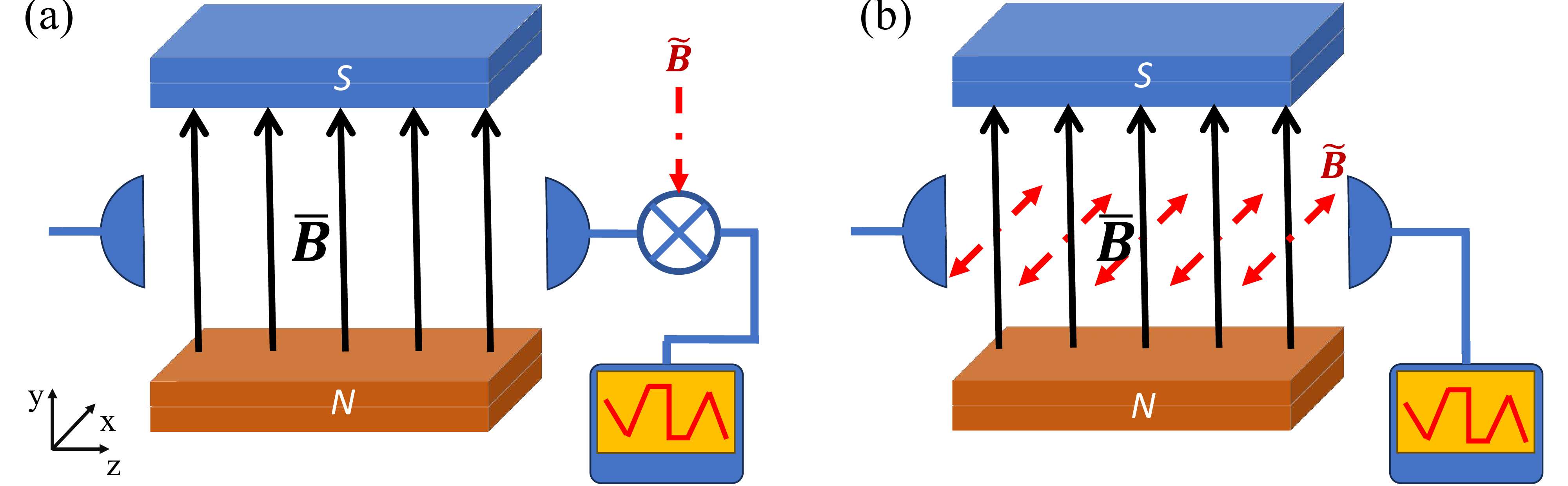}
\caption{Electromagnetic response detections of axions with a one-dimensional cavity biased by a high stationary magnetic field $\bar{B}$: (a) The generated EMR signal of the passing axion is coherently amplified ex-situ by applying an additional RF-field; and (b) the EMR signal of the axion is significantly enhanced in-situ,  due to the interaction between the excited magnetic mode and the passing axion field.\label{fig1}}
\end{figure}

Roughly, there are three approaches to detect the axions. The first one is the gravitational probes, i.e., searching for the possible signatures in CMB and Pulsar Timing Arrays (PTA)~\cite{ARNPS2015,2021AxionSA,PRL20241}, for the detection of the axions with exceptionally light masses. 
Next, the CASPEr-type experiments ~\cite{PRX2014,PRDFGC2013} and PandaX-like experiments~\cite{PRL20172} have been used to probe the observable effects induced by the axion-nucleon and axion-electron interactions for the axions with the slightly larger masses. 
Thirdly, it is believed that the Haloscope-type detectors (HTDs), shown schematically in Figure~\ref{fig1}(a) without the radio-frequency (RF) magnetic field $\tilde{B}(t)$ and based on the inverse Primakoff effect~\cite{PR1951}, are particularly suitable to detect the electromagnetic responses (EMRs) of the axions with $\rm \mu eV$-order masses. However, the results obtained by using all the existing HTDs biased only by the high stationary magnetic fields (SMFs), typically including the ADMX~\cite{PRL20201,RSI2021} and CASTRBF~\cite{RSI2021,ARNPS2015,PRL19871}, Optical Ring Cavity Technology~\cite{PRL20181},
TOORAD~\cite{PRL20191},
HAYSTAC~\cite{NIMPRSA2017},
CAPP~\cite{PRL20211},
ORGAN~\cite{PDU2017}, and TASEH~\cite{PRL20222} are still null. This implies that the detection sensitivity of these existing HTDs are still expected to be further improved.

Basically, the dark matter axion could be treated as a pseudo-scalar particle in QCD, and its interaction with the electromagnetic field can be described by the following Lagrangian density~\cite{NPB2024}:
\begin{align}\label{eq1}
\mathscr{L} = {F}_{\mu\nu} {F}^{\mu\nu} +j_\mu A^\mu                        
-\frac{1}{2}m_a^2a^2                         
+ \frac{1}{2}\partial_\mu a \partial^\mu a        
-g_{a\gamma\gamma} a {F}_{\mu\nu} \tilde{{F}}^{\mu\nu},   
\end{align}	
here, $j_\mu$ denotes the usual four-current density, $ A^\mu$ stands for the electromagnetic four-potential, ${F}_{\mu\nu}$ and $\tilde{{F}}_{\mu\nu} $ are the electromagnetic field tensor and its dual, respectively. 
Here, $a$ is the axion field with the mass $m_a$ and the oscillation frequency $\omega_a=m_ac^2/\hbar$. Here, $c$ is the speed of light and $\hbar$ the reduced Planck constant. Usually, the decay parameter $f_a$ of the axion can be described as 
$f_a=10^{12}{\rm GeV}$, and $g_{a\gamma\gamma}=g_\gamma f_a^{-1}\alpha/\pi$
is the axion-photon coupling strength, with $\alpha$ being the fine-structure constant and $g_\gamma$ a dimensionless constant depending on the axion models. Specifically, $g_\gamma=-0.97$ in the KSVZ model~\cite{A2023,SS2022},
and $g_{\gamma}=0.36$ in the DFSZ one \cite{PRB1981,SJNP1980}.
Generically, the achievable minimum value of $g_{a\gamma\gamma}$ refers to the detection sensitivity of the axion field. 

In principle, by further enhancing the strength of the applied SMF, the detection sensitivity of the existing HTD can be further improved. However, its feasibility is limited both technologically and financially. Therefore, an improved approach shown schematically in Figure~\ref{fig1}(a) could be utilized to 'ex-situ' coherently amplify the EMR signal $S_{a\rightarrow \gamma}^{\rm C}(t)$ generated in the usual HTDs~\cite{2021Probing,PRD2019,PRD2023,2018Interference}, by coherently mixing the additionally applied RF-frequency signal $\tilde{B}(t)$ with the original signal $S_{a\rightarrow\gamma}^{\rm C}(t)$. This approach has certainly improved the detectability of $S_{a\rightarrow\gamma}^{\rm C}(t)$, which can be coherently amplified as $P_{\rm I}(t)$ now. But additional noise is introduced simultaneously by the application of $\tilde{B}(t)$. Consequently, the achievable detection sensitivity remains still very limited, at least theoretically.

Alternatively, in the present work we propose a conceptually novel approach, see Figure~\ref{fig1}(b) for simple comparison, to upgrade the existing HTD configuration (hereafter referred to as UHTD), i.e., keeping the longitudinal high SMF unchanged but additionally just apply a weak transverse RF-excited magnetic field $\tilde{B}(t)$ to excite the cavity's magnetic resonant mode for in-situ enhancing the axion-photon interference energy.
As a consequence, the power of the generated EMR signal $S_{\rm rf}^{(1)}(t)$ is now proportional to the axion-photon coupling strength, unlike in the existing HTD wherein the EMR signal $S_{a\rightarrow\gamma}^{\rm C}(t)$ is just proportional to the square of the axion-photon coupling strength. 
  
Given the transverse RF-excited field technique has
been widely applied in the magnetic resonance imagings,  various filtering- and heat conduction techniques, it is argued that the achievable detection sensitivity of the UHTD for the EMR detection of axion field in RF-band could be enhanced by {\color{blue}$0.3\sim 1$} orders of magnitude compared with the existing HTDs. Therefore, it is expected that the proposed UHTD might greatly push the progress for the Haloscope detection of axion field in RF-band. 

The paper is arranged as follows. In Sec.~2 we first review briefly how the axion fields can be detected with the existing HTDs and then discuss the limitation in the previous approaches by additionally applying the RF signals to ex-situ coherently amplify the original EMR signals for improving their sensitivity.  In Sec.~3, we demonstrate how the axion-photon interference energy can be significantly enhanced in the UHTD. In Sec.~4, we specifically investigate the achievable detection sensitivity by probing the first-order electromagnetic response signal generated in the UHTD, and we evaluate the noise effective power of the UHTD through a rigorous noise analysis. Finally, in Sec.~5, we summarize our work and discuss the feasibility of the proposed UHTD.

\section{The signals generated in the HTDs for axion fields detection} 

In this section we first review briefly how the axion fields can be detected with the existing HTDs and then discuss the limitation in the previous approaches by additionally applying the rf signals to ex-situ coherently amplify the original EMR signals for improving their sensitivity. 
It is well-known that, in the usual HTD, which is biased by a high stationary magnetic field ${\bm B}^{(0)}$, the axion-modified Maxwell equation~\cite{NPB2024}:
\begin{equation}\label{eq2}
\left\{
\begin{aligned}		   
&\nabla\cdot\bm{E}(t)=\frac{1}{\varepsilon_0}g_{a\gamma\gamma}\bm{B}^{(0)}\cdot\nabla a,\\[3pt]
&\nabla\times\bm{B}(t)-\frac{1}{c^2}\partial_t\bm{E}(t)=-\mu_0g_{a\gamma\gamma}\bm{B}^{(0)}\partial_ta,\\
\end{aligned}
\right.
\end{equation}
can be obtained from the Lagrangian Eq.~\eqref{eq1}. Here, $a=a({\bm{x}},t)$ is the axion field, $ \varepsilon_0$ is permittivity and $\mu_0$ the permeability in vacuum; ${\bm B}(t)$ and ${\bm E}(t)$ are the magnetic- and electric field densities of the EMR signal generated in the HTD, respectively.
Without the axion field and neglecting the background EM noise, the  electromagnetic field in the HTD cavity can be expressed as $E(t=0)=0$ and $B(t=0)=B^{(0)}$. Now, if an axion field passes through the HTD's cavity, an EMR signal with $E(t>0)\neq 0$ and $B(t>0)>B^{(0)}$ can be generated for the detection.  

Specifically, when an approximately isotropic axion wave, i.e.,  $\nabla a\ll\partial_ta$ and thus
$a(\bm{x},t)=\mathrm{Re}(a_0e^{-i(\omega_at+\phi_a)})$ (with $\phi_a$
 being the random phase within the range 
$[0,2\pi)$~\cite{2018axionphase,QiaoliYang}) and $\rho_{\rm eff}\sim 0$, passes through a cavity biased by a
high SMF  
$\bm {B}^{(0)}=\bar{B}(\bm{x})$ (which is assumed to be along the $y$-direction),
then an effective current:
\begin{align}\label{eq12}
	\bm{j}_{\rm eff}(\bm{x},t)=-g_{a\gamma\gamma}\bar{B}(\bm{x})\partial_ta,
\end{align}
 it satisfies the following dynamical equation:
\begin{align}\label{eq11}	
(\frac{d^2}{dt^2}+\Gamma_l\frac{d}{dt}+\omega_l^2)\psi_l(t)=\mu_0\bm{j}_{\rm eff}(\bm{x},t),
\end{align}
where $\Gamma_l=\omega_l/Q_l$
is the energy dissipation rate of the $l$-th mode with $Q_l=\omega_l/\Gamma_l$ being its quality factor.
By solving \eqref{eq11}, the response electromagnetic field induced by the axion can be obtained. 
Above, we have assumed that the response is resonant for simplicity, i.e., $\omega_a=\omega_l$, and the  experimental parameters are set typically as~\cite{SJNP1980}:
$g_\gamma=0.36$, $\rho_a=2^{-1}\times10^{-24}$ $\rm g/cm^3$,	
$\bar{B} =8$ T, $ V=1$~$\rm m^3$, $Q_l=10^4$, $\omega_a=2\pi\times 1$~GHz, and $f_am_a=6\times10^{15}\rm ~(eV)^2$, the response electric field is $|\bar{E}_y^{(1)}|=1.0095\times10^{-10}\rm (V/m)$, and the magnetic field is $|\bar{B}_x^{(1)}|=3.365\times10^{-17}\rm T$.
    Obviously, both the electric- and magnetic field intensities of the response signals of the passing axion field, generated in the usual HTD, are virtually undetectable; even with the most sensitive electric
~\cite{NC2024,RPP2023}
    and magnetic field detectors~\cite{ARBE2007} available today, they are  still a few orders of magnitude away from the sensitivities required to effectively probe these significantly weak signals. Therefore, the energy detection of the electromagnetic response signal of the passing axion field is necessary. 
The detectable power of the axion response signal in the HTDs, for the resonant response with $\omega_l=\omega_a$, can be expressed as: 
\begin{align}\label{eq2.0}
P^{(2)}_{a\rightarrow\gamma}
=g_{a\gamma\gamma}^2\times\frac{2\rho_a\omega_a }{m_a^2\mu_0}\bar{B}^2Q_lC_lV,
\end{align}
with $C_l=(\int_Vd^3x\bar{B}(\bm{x})\cdot\bm{e}_l(\bm{x}))^2/(\bar{B} ^2V)$ being the form factor of the $l$-th mode of the electromagnetic field. The relevant derivation is provided in Appendix A.
Currently, most axion detection devices are based on this theory, in which the detectable signal is proportional to the square of the extremely small coupling parameter $g_{a\gamma\gamma}$. This explains why all experimental searches using HTDs have so far yielded no positive results. Upgrading the usual HTD-type installations to further improve their achievable detection sensitivities, for capturing the weaker electromagnetic response signals of the axion fields, is particularly desirable. 

Physically, to enhance the detectabilityof axion's electromagnetic response, one must either continue to increase the strength of the steady-state strong magnetic field $\bar{B}$ for generating the stronger electromagnetic response signal (as Eq.~\eqref{eq2.0} shows that the power of such an EMRs is proportional to increase the strength of SMF $\bar{B}$), or coherently amplify the amplitude of the generated electromagnetic response signal for improving its detectability.
To overcome the limitations of the former approach, which is constrained by the high manufacturing cost of the steady-state magnetic field, Ref.~\cite{PRD2023} proposed a novel method: injecting photons into the system and using a power detector to measure the variance of the interference signal from axions.
However, the detected signal in Ref.~\cite{PRD2023} remains the traditional HTDs second-order axion power signal $P_{\rm rf}^{(2)}\propto g_{a\gamma\gamma}^2$. By injecting intense photons, the weak axion signal is expected to be amplified as measurable variance $\hat{\sigma}^2(N_T) = N_s+{\rm others}$ (where $N_s \propto g_{a\gamma\gamma}^2$), and the axion response signal power should be greater than $10^{-6}$ mW. However, for extremely weak axion-photon coupling, generating such a strong axion signal remains highly challenging (see Appendix B for details)~\cite{PRD2023}.

\section{The EM response of the axion field passing through the proposed UHTD}

In this section, we demonstrate how the axion-photon energy transferring rate can be significantly enhanced in the  UHTD, whose principle has been described briefly in Sec.~1. Differing from the external coherent amplification of the responded signal mentioned above, which is merely aimed at improving its detectability, the additionally applied RF- field applied here is used to excite the cavity's magnetic resonant mode  to significantly enhancing the axion-photon converted rate for generating the significantly detectable signals.

Now, let us consider a simplified configuration, shown in Figure~\ref{fig1}(b), to upgrade the existing HTDs for the sensitive EMR detection of the axion. 
Besides the 
usual high SMF $\bar{B}$ applied along the $y$-direction, a RF-excited field is applied here to excite the cavity's magnetic resonant mode along the $x$ - direction with the amplitude being $\tilde{B}\ll \bar{B}$.  As a consequence, the background magnetic field in the existing HTD can be upgraded as:
\begin{align}\label{eq16}
	\bm{B}^{(0)}(\bm{x},t)=\bar{\bm{B} }\bm{e}_y+\tilde{B}\mathrm{Re}\left(e^{i(\bm{k}_B\cdot\bm{x}-\omega_{B}t)}\right)\bm{e}_x,
\end{align}
where $\tilde{B}$, $\bm{k}_B$ and $\omega_B$ is the amplitude, wave vector and frequency of the RF-excited field, respectively. Taking the background magnetic field Eq.~\eqref{eq16} into the axion-modified Maxwell equations shown in Eq.~\eqref{eq2}, the equations solution can be formally expressed as:
\begin{equation}\label{eq17}
    \left\{
\begin{aligned}
	\bm{B}=&\bm{B}^{(0)}+\bm{B}^{(1)}+\mathcal{O}^{(2)}(g_{a\gamma\gamma}),\\[3pt]
	\bm{E}=&\bm{E}^{(0)}+\bm{E}^{(1)}+\mathcal{O}^{(2)}(g_{a\gamma\gamma}),
\end{aligned}
\right.
\end{equation}
the superscripts: $(0)$ and $(1)$, indicate the effects related to the zeroth- and first-order (1st) responses of the axion-photon coupling, and the effects related to the second-order (2nd) and higher ones, i.e., $\mathcal{O}^{(2)}$, are safely neglected. 
Given the zeroth-order effect can be easily calculated by solving the conventional Maxwell equation in the absence of axion-photon coupling. The 1st EMR of the axion can be described by
\begin{equation}\label{eq20}
\renewcommand{\arraystretch}{1.5} 
    \left\{
\begin{aligned}	&\nabla\cdot\bm{E}^{(1)}=\frac{1}{\varepsilon_0}g_{a\gamma\gamma}\bm{B}^{(0)}\cdot\nabla a\sim 0,\\[3pt]
		&\nabla\bm{B}^{(1)}-\frac{1}{c^2}\partial_t\bm{E}^{(1)}=\mu_0\bm{j}^{(1)}_{\rm eff}(\bm{x},t),
\end{aligned}
\right.
\end{equation}
for the approximately isotropic axion wave, i.e., the axion field can be described as $a(\bm{x},t)\approx Re(a_0e^{-i(\omega_at+\phi_a)})$. 

Under the Lorenz gauge condition; where $\bm{B}^{(1)}=\nabla\times\bm{A}^{(1)},\bm{E}^{(1)}=-\partial_t\bm{A}^{(1)}-\nabla \phi^{(1)}$, $\nabla\cdot \bm{A}^{(1)}+\partial_t\phi^{(1)}=0$, we can conclude that:
$   -\nabla^2\bm{A}^{(1)}+\partial_t^2\bm{A}^{(1)}/c^2=\bm{j}_{\rm eff}^{(1)},
$
When a RF-excited magnetic field is added perpendicular to the direction of the steady magnetic field, the resulting axion response effective current will become
\begin{align}
    \bm{j}_{\mathrm{eff}}^{(1)}=-\frac{1}{4c}g_{a\gamma\gamma}[\bar{B}\bm{e}_{y}+\tilde{B}\operatorname{Re}(e^{i(\bm{k}_{B}\cdot\bm{x}-\omega_{B}t)})\bm{e}_{x}]\times\partial_{t} a.
\end{align}

\begin{figure}[ht]
    \centering
    \includegraphics[width=0.9\linewidth]{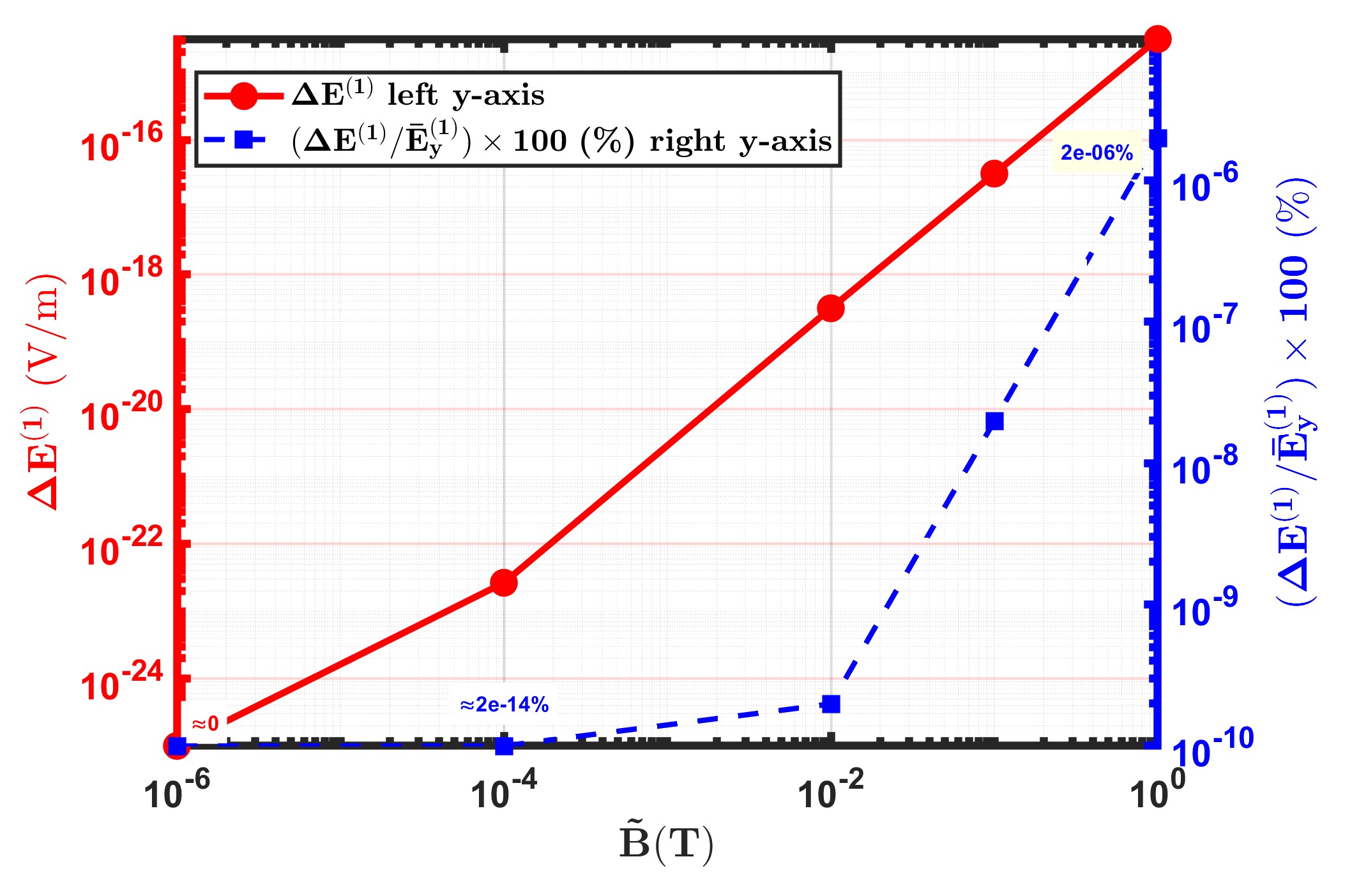}
    \caption{The increment and relative percentage increment of the 1st-order axion-induced electric field induced by the RF-excited magnetic field.}
    \label{fig2}
\end{figure}

In the cavity, the electromagnetic vector potential $\bm{A}^{(1)}$ can be expanded in terms of a series of discrete modes: $\bm{A}^{(1)}=\sum_l\bm{u}_l(\bm{x})\psi_l(t)$. For the $l$-th cavity mode with the dissipation $\Gamma_l$, its response to the axion field, i.e., the signal transported along the $x$-direction can be divided into space part and time-dependent part. And the space part satisfies the orthogonality, normalization, and completeness relations. Then, the time-dependent part will satisfy the following equation:
\begin{align}\label{eq22}
   & (\frac{d^2}{dt^2}+\Gamma_l\frac{d}{dt}+\omega_l^2) 
\psi^{(1)}_{x,l}(t)
    =a_{lx}\Omega\mathrm{Re}(e^{-i(\omega_at+\phi_a)})\mathrm{Re}(e^{-i\omega_Bt}), 
\end{align}
with $\Omega=
cg_{a\gamma\gamma}\omega_a|a_0|/4$ and $a_{lx}=\int_V\tilde{B}(\bm{x})\cdot\bm{u}_l^*(\bm{x})d^3x $ is the overlap coefficient of the effective flow to cavity mode.  The electromagnetic vector potential $x$-component solution reads 
$\bm{A}^{(1)}_{x,l}(\bm{x},t)=\left(A^{(1)}_{x,l^+}(\bm{x},t)+A^{(1)}_{x,l^-}(\bm{x},t)\right)\bm{e}_x$, where
$\bm{A}^{(1)}_{x,l^\pm}(\bm{x},t)=\mathrm{Re}[{ia_{lx}\Omega \bm{u}_l(\bm{x})e^{-i(\Delta_{aB}^\pm t+\phi_a)}}/\\(\omega_{l}^2-(\Delta_{aB}^\pm)^2-i\Gamma_{l}(\Delta_{aB}^\pm))]$, 
and $\Delta^\pm_{aB}=\omega_a\pm\omega_B$.
Similarly, for the response along the $y$-direction, we have
\begin{align}\label{eq25}
   & (\frac{d^2}{dt^2}+\Gamma_l\frac{d}{dt}+\omega_l^2) 
\psi^{(1)}_{y,l}(t)
    =a_{ly}\Omega\mathrm{Re}(e^{-i(\omega_at+\phi_a)}),
\end{align}
where, $a_{ly}=\int_V\bar{B}(\bm{x})\cdot\bm{u}_l^*(\bm{x})d^3x $, whose solution reads
$\bm{A}^{(1)}_{y,l}(\bm{x},t)=a_{ly}\Omega\bm{u}_{l}(\bm{x})
\mathrm{Re}[ie^{-i(\omega_at+\phi_a)}/\\(\omega_{l}^2-\omega_a^2-i\Gamma_{l} \omega_a))]\bm{e}_y$. According to the relationship between electromagnetic field and the magnetic vector potential, the electromagnetic Eq.~\eqref{eq17} in the cavity can be specifically expressed as:
\begin{equation}\label{eq26}
    \left\{
\begin{aligned}
	\bm{E} =& \tilde{E}_x^{(1)}\bm{e}_x +\left[E_y^{(0)} +\bar{E}_y^{(1)}\right]\bm{e}_y,\\[3pt]
	\bm{B}=& \left[\tilde{B}_x^{(0)} 
                        +\bar{B}_x^{(1)} \right]\bm{e}_x  + \left[\bar{B} + \tilde{B}_y^{(1)}\right]\bm{e}_y,
\end{aligned}
    \right.
\end{equation}
with
\begin{equation}
    \left\{
\begin{aligned}\label{eq19}
	&\bm{E}_y^{(0)}(\bm{x},t)=-\mathrm{Re}(c\tilde{B}e^{i(\bm{k}_B\cdot\bm{x}-\omega_Bt)})\bm{e}_y,\\[3pt]
&
 \bm{B}^{(0)}(\bm{x},t)=\bar{\bm{B} }\bm{e_y}+\mathrm{Re}(\tilde{B}e^{i(\bm{k_B}\cdot\bm{x}-\omega_{B}t)})\bm{e}_x,
\end{aligned}
\right.
\end{equation}
for the zeroth-order EMR signal,
\begin{equation}\label{eq27}
    \left\{
\begin{aligned}
		\tilde{E}^{(1)}_x(\bm{x},t)&=
a_{lx}\Omega\bm{u}_l(\bm{x})(\Delta_{aB}^+ \Phi_{l+}+\Delta_{aB}^-\Phi_{l-}),\\[3pt]
		\bar{E}^{(1)}_y(\bm{x},t)&=\omega_aa_{ly}\Omega\bm{u}_l(\bm{x})\Phi_l,
\end{aligned}
 \right.
\end{equation}
and
\begin{equation}\label{eq28}
    \left\{
\begin{aligned}
	\tilde{B}^{(1)}_y(\bm{x},t) = &-\bm{k}_la_{lx}\Omega\bm{u}_l(\bm{x})[ 
   \Phi_{l+}
    +\Phi_{l-}],\\[3pt]
		\bar{B}^{(1)}_x(\bm{x},t)=&-\bm{k}_la_{ly}
    \Omega\bm{u}_l(\bm{x})\Phi_l,
\end{aligned}
   \right.
\end{equation}
for the 1st EMR signal. Above,  $\Phi_{l\pm}=\mathrm{Re}\left(e^{-i(\Delta_{aB}^\pm t+\phi_a)}/[\omega_{l}^2-(\Delta_{aB}^\pm)^2-i\Gamma_{l}(\Delta_{aB}^\pm)]\right)$, $\Phi_{l}=\mathrm{Re}\left(e^{-i(\omega_at+\phi_a)}/[\omega_{l}^2-\omega_a^2-i\Gamma_{l}\omega_a]\right)$, and $\bm{k}_l=\omega_l/c$.
Obviously, these intensities are proportional to the weak axion-photon coupling strength $g_{a\gamma\gamma}$. 
Alongside the original Haloscope response fields $\bar{E}_y^{(1)}$ and $\bar{B}_x^{(1)}$, the RF-excited magnetic field induces an additional first-order axion-converted response, $\tilde{E}_x^{(1)}$ and $\tilde{B}_y^{(1)}$. Consequently, the axion-response electric field is enhanced from $|\bar{E}_y^{(1)}|$ to $(|\bar{E}_y^{(1)}|^2 + |\tilde{E}_x^{(1)}|^2) ^{1/2}$. {\color{blue}Concomitantly, from Eq.~\eqref{eq26}, the cavity energy flux density can be expressed order by order in the axion coupling constant $\bm{S}_{\text{rf}}=\sum_{i=0}^2\bm{S}_{\text{rf}}^{(i)}(t,g_{a\gamma\gamma}^i)$.} In Figure~\ref{fig2}, we shows schematically such a weak contribution from the axion-photon energy conversion brought by the transverse RF-excited field.
While, with such a RF-excited magnetic field driving, the 1st interference energy flux density of the axion response signal can be additionally generated as:
\begin{align}\label{eq29}
    \bm{S}^{(1)}_{\rm rf}&=\frac{1}{\mu_0}\left(\tilde{E}_x^{(1)}\bar{B} -E_y^{(0)}\bar{B}_x^{(1)}-\bar{E}_y^{(1)}B_x^{(0)}\right)\bm{e}_z\nonumber\\
        &=-\frac{1}{\mu_0}a_{lx}\Omega\bar{B}\bm{u}_l(\bm{x})[\Delta_{aB}^+ \Phi_{l+}+\Delta_{aB}^-\Phi_{l-}]\bm{e}_z+\frac{\omega_l}{\mu_0}a_{ly}\Omega\bm{u}_l(\bm{x})\tilde{B}\Phi_l\mathrm{Re}\left(e^{i(\bm{k}_B\cdot\bm{x}-\omega_Bt)}\right)\bm{e}_z,
\end{align}
besides the zeroth-order energy flux, $\bm{S}^{(0)}_{\rm rf}=E_y^{(0)}B_x^{(0)}\bm{e}_z/\mu_0$ (shown in Eq.~\eqref{eq19}), which is independent of axion. 
 
Continuously, the 2nd energy flux density of the present EMR signal of the axion field can be calculated as:
\begin{align}\label{37}
		\bm{S}^{(2)}_{\rm rf}&=\frac{\Gamma_l}{\mu_0}\bm{E}^{(1)}\times\bm{B}^{(1)}=\frac{\Gamma_l}{\mu_0}\left(\tilde{E}_x^{(1)}\tilde{B}_y^{(1)}-\bar{E}_y^{(1)}\bar{B}_x^{(1)}\right)
		=\frac{\Gamma_l}{\mu_0}\left(\partial_t A^{(1)}_{x,l}\partial_z A^{(1)}_{x,l}+\partial_tA^{(1)}_{y,l}\partial_zA^{(1)}_{y,l}\right)\nonumber\\
		=&-\frac{\Gamma_l}{\mu_0}\bm{k}_la_{lx}^2\Omega^2\bm{u}_l^2(\bm{x})\left[(\Delta_{aB}^+\Phi_++\Delta_{aB}^-\Phi_{l-})(\Phi_++\Phi_{l-})\right]+\frac{\Gamma_l}{\mu_0}\bm{k}_l\omega_aa_{ly}^2\Omega^2\bm{u}_l^2(\bm{x})\Phi_l^2,
\end{align}
and its time-average reads
\begin{align}\label{36}
\langle\bm{S}^{(2)}_{\rm rf}\rangle=cg_{a\gamma\gamma}^2|a_0|^2\omega_aQ_l(\bar{B}^2+\tilde{B}^2) C_lV\bm{e}_z,
\end{align}
which is obviously  proportional to the square of axion-photon coupling strength (as $\Omega=
cg_{a\gamma\gamma}\omega_a|a_0|/4$ defined in Eq.~\eqref{eq22}), the
squares of amplitudes of the biased high SMF (as $a_{ly}\sim\bar{B} $ shown in Eq.~\eqref{eq25}), and the additionally applied RF-excited field (as $a_{lx}\sim\tilde{B}$ shown in Eq.~\eqref{eq22}. Since $\bar{B}\gg\tilde{B}$, the 2nd energy response is mainly contributed by the biased high SMF. In contrast, the 1st interference energy flux density response of the passing axion field is only generated by the applied RF-excited field, i.e., for $\tilde{B}=0$ and thus $a_{lx}=0$, we have  $S^{(1)}_{\rm rf}=0$.

In principle, the phase factor might be treated as an arbitrarily unknown random variable, and thus if it influences the axion-photon energy conversion should be discussed. 
Following Refs.~\cite{2018axionphase,QiaoliYang},  the 1st interference energy flux density of the axion's response signal, shown in Eq.~\eqref{eq29} for $\phi_a=0$, should be replaced as the $\phi_a$-dependent one $S_{\rm rf}^{(1)}(\phi_a)$.   
Consequently, under the resonant driving, i.e., 
$\omega_a=\omega_B=\omega_l$, the time-average of $S_{\rm rf}^{(1)}$  should be expressed as ~\cite{2018axionphase,QiaoliYang}, \begin{align}\label{C6}
     \langle \bm{S}_{\rm rf}^{(1)}(\phi_a)\rangle&=\lim_{T\rightarrow\infty}\frac{1}{T}\int_0^T\frac{\bm{e}_z}{\mu_0}\left(\tilde{E}_x^{(1)}\bar{B} -E_y^{(0)}\bar{B}_x^{(1)}-\bar{E}_y^{(1)}B_x^{(0)}\right)dt\nonumber\\
&=cg_{a\gamma\gamma}|a_0|\bar{B}\tilde{B}Q_l C_lV\cos{(\phi_a)}\bm{e}_z.
 \end{align}
 where  $C_l=(\int_Vd^3x\bar{B}(\bm{x})\cdot\bm{u}_l(\bm{x})\int_Vd^3x\tilde{B}(\bm{x})\cdot\bm{u}_l(\bm{x}))/(\bar{B} \tilde{B}V)$ is the form factor of the $l$-th mode electromagnetic field in the cavity.
By making the statistical average on the random phase, we have \begin{align}\label{C7}
     \overline{\langle \bm{S}_{\rm rf}^{(1)}(\phi_a)\rangle} =cg_{a\gamma\gamma}|a_0|\bar{B}\tilde{B}Q_l C_lV\overline{\cos(\phi_a)}\bm{e}_z,
 \end{align}
 where $\overline{\cos(\phi_a)}$ is the statistical average of $\cos(\phi_a)$ for the unknown stochastic phase $\phi_a\sim [0,2\pi)$~\cite{2018axionphase,QiaoliYang}. 
 As a result, it cannot be directly detected, as $\overline{\cos(\phi_a)}$ is zero for the long-time averages.
 
 Alternatively, we show below that the stable output of 1st-order axion response signals can be implemented by using the usual IQ-mixer  modulation technique, although the phase of the 1st-order axion response signal is still random~\cite{IQ1,IQ2,IQ3}. In fact, the 1st interference energy flux density shown in Eq.~\eqref{eq29} can be rewritten as:
 \begin{align}\label{St}
    S^{(1)}_{\rm rf}(t)=|S_{\rm rf}^{(1)}|[\cos(\Delta_{aB}^+t+\phi_a)+\cos(\Delta_{aB}^-t+\phi_a)],
\end{align}
with $|S_{\rm rf}^{(1)}|=ca_0g_{a\gamma\gamma}\omega_a^2\bar{B}\tilde{B}/(4\mu_0\Gamma\omega_a)$ being its amplitude.
Under the resonant condition, i.e., 
$\omega_a=\omega_B=\omega_l$, we have
$
S^{(1)}(t)\sim |S_{\rm rf}^{(1)}|[\cos(2\omega_Bt+\phi_a)+\cos{(\phi_a)}]$, which can be split into two signals $S^{(1)}_1(t)$ and $S^{(1)}_2(t)$ with the same amplitude of $|S_{\rm rf}^{(1)}|/2$. They are mixed with the local coherent electromagnetic signals  $L_1(t)\sim L\cos(2\omega_Bt)/2$ and $L_2(t)\sim L\sin(2\omega_Bt)/2$, where $L$ is the known amplitude. After the low-pass filtering, we get the $I(\phi_a)$-channel signal:
$    I(\phi_a)=\frac{1}{4}|S_{\rm rf}^{(1)}|L\cos(\phi_a),
$
and the $Q(\phi_a)$-channel signal:
$
    Q(\phi_a)=-\frac{1}{4}|S_{\rm rf}^{(1)}|L\sin(\phi_a),
$
respectively. 
Finally, the amplitude of the 1st interference
energy flux density amplitude $|S_{\rm rf}^{(1)}|$, can be determined by the demodulation signal with the usual IQ mixer technique, i.e.,
\begin{align}\label{S1}
D= \sqrt{I^2(\phi_a)+Q^2(\phi_a)}=\frac{1}{4}|S_{\rm rf}^{(1)}|L,
\end{align}
which is proportional to $g_{a\gamma\gamma}$ and independent of the signal phase (see Appendix C for details). {\color{blue}We can recover the amplitude of $|S_{\text{rf}}^{(1)}|$ from the output values of the IQ quadrature mixer.} The next question is, if such a first-order power signal could be detected at a sufficiently sensitive level.

\section{The achievable detection sensitivity.}

{\color{blue}
After the interference mixing between the axion-induced first-order response electric field $\bar{E}_y^{(1)}$ and the background driving field $E_y^{(0)}$, when the detector normal is parallel to the direction of the energy flux density, the signal power received by the detector with an effective detection area of $A=1\,\text{mm}^2$ is $\sum_{i=0}^2P_{\text{rf}}^{(i)}=\sum_{i=0}^2S_{\text{rf}}^{(i)}A$, where $P_{\text{rf}}^{(0)}\propto (E_y^{(0)})^2$ is the local oscillator signal power, $P_{\text{rf}}^{(2)}\propto (\bar{E}_y^{(1)})^2$ is the axion signal power, and $P_{\text{rf}}^{(1)}=\sqrt{2P_{\text{rf}}^{(0)}P_{\text{rf}}^{(2)}}\propto E_y^{(0)}\bar{E}_y^{(1)}$ is the first-order signal power. Since the signal power is much smaller than the local oscillator power, i.e., $P_{\text{rf}}^{(2)}\ll P_{\text{rf}}^{(0)}$, the second-order power can be neglected.
The noise in this process, $P_N= P_{\text{shot1}} +P_{T}$, is mainly affected by shot noise $P_{\text{shot1}}$ and thermal noise $P_T$. Here, the shot noise includes the contributions from the local oscillator power, the axion second-order power, the background noise power $P_B \sim 10^{-17}\,\text{W}$, and the dark count power $P_D \sim 10^{-20}\,\text{W}$ at the detection terminal within the detection bandwidth $\Delta f$:
\begin{align}
    P_{\rm shot1}=\sqrt{(P_{\text{rf}}^{(0)}+P_{\text{rf}}^{(2)}+P_B+P_D)\hbar\omega_a/\tau}.
\end{align}
The thermal noise is:
\begin{align}
   P_{T}=\hbar\omega_a\left(\frac{1}{2}+\frac{1}{e^{\hbar\omega_a/k_BT_{\rm sys}}-1}\right)\Delta f,
\end{align}
where $k_B$ is the Boltzmann constant, and $T_{\rm sys}$ is the system temperature
Under the influence of the total noise, the signal-to-noise ratio (SNR) of this upgraded axion detection scheme can be defined as~\cite{SNR1}:
\begin{align}
    \text{SNR}=\frac{P_{\text{rf}}^{(1)}\sqrt{t_{in}\Delta f}}{P_N}=\frac{\sqrt{2P_{\text{rf}}^{(0)}P_{\text{rf}}^{(2)}}\sqrt{t_{in}\Delta f}}{\sqrt{(P_{\text{rf}}^{(0)}+P_{\text{rf}}^{(2)}+P_B+P_D)\hbar\omega_a/\tau}+\hbar\omega_a\left(\frac{1}{2}+\frac{1}{e^{\hbar\omega_a/k_BT_{\rm sys}}-1}\right)\Delta f},
\end{align}
where $t_{in}$ is the  integration time. Compared with the shot noise $P_{\text{N0}}\sim 10^{-11}\,\text{W}$ arising from the strong local oscillator power $P_{\text{rf}}^{(0)}$ under the excited field amplitude of $\tilde{B}=10^{-6}\,\text{T}$, the noises induced by the second-order power $P_{\text{rf}}^{(2)}$, background noise $10^{-19}\,\text{W}$, dark count noise $10^{-20}\,\text{W}$, and thermal noise ($300\,\text{K}, P_T\sim 10^{-20}\,\text{W}$) are extremely small and negligible. Therefore, the SNR can be expressed as:
\begin{align}
    \text{SNR}=\frac{\sqrt{2P_{\text{rf}}^{(0)}P_{\text{rf}}^{(2)}}\sqrt{t_{in}\Delta f}}{\sqrt{P_{\text{rf}}^{(0)}\hbar\omega_a/\tau}}=\sqrt{\frac{2P_{\text{rf}}^{(2)}t_{in}\Delta f}{\hbar\omega_a/\tau}}.
\end{align}
\begin{figure}
    \centering
    \includegraphics[width=0.8\linewidth]{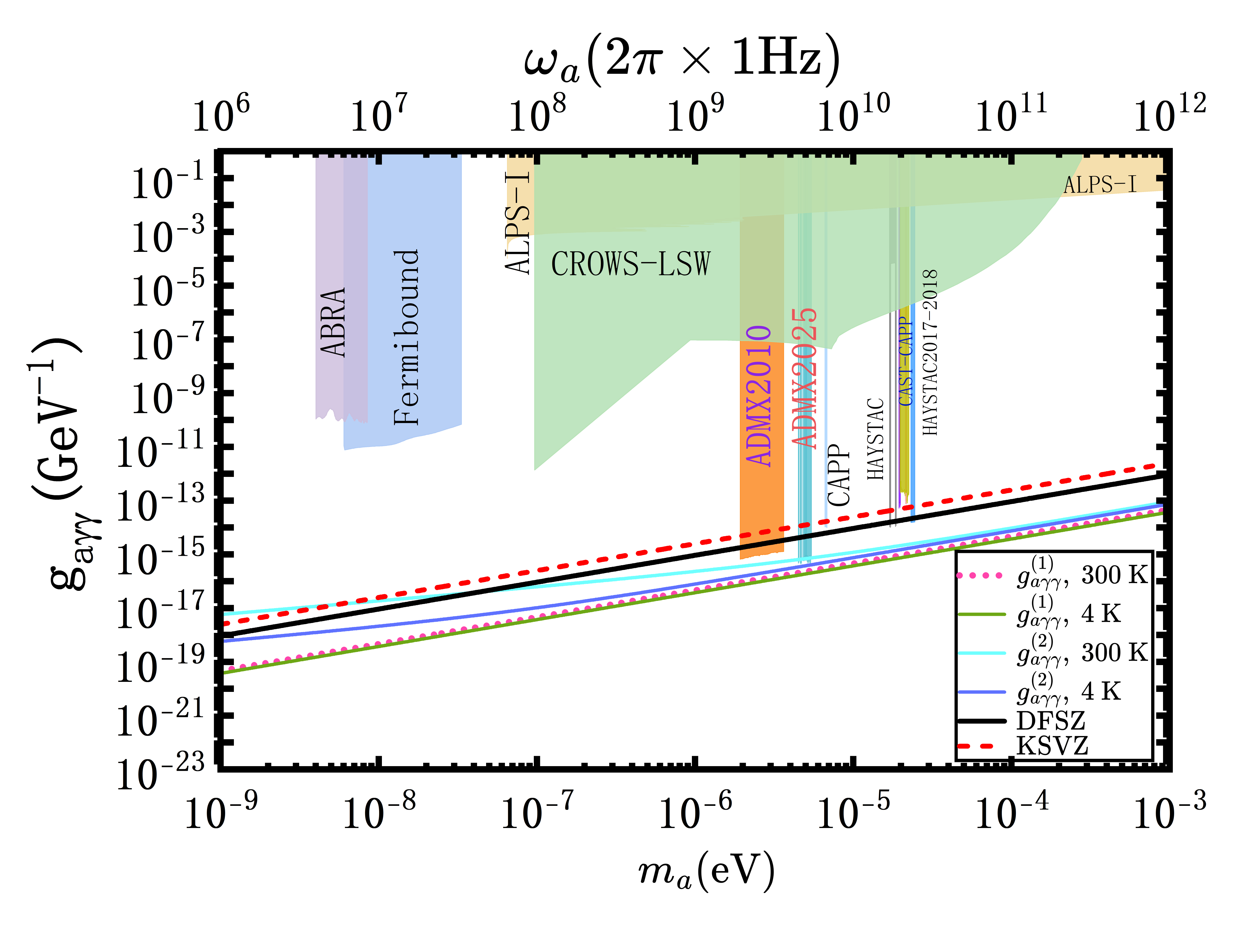}
    \caption{Sensitivity comparison of axion electromagnetic response detection of different masses in microwave cavity with strong magnetic field. In the figure, $g_{a\gamma\gamma}^{(1)}$ and $g_{a\gamma\gamma}^{(2)}$ represent the corresponding sensitivity parameters of the first order (with RF field driven) and second order (without RF field drive) energy response signal detection, respectively. Black and red represent the expected sensitivity of the DFSZ model and the KSVZ model, respectively; The area covered by various colors indicates the sensitivity achieved by the corresponding experiment. At 4 K, the quality factor of the resonant cavity is $Q = 7.5 \times 10^4$, and at 300 K, $Q = 5 \times 10^4$.}
    \label{fig3}
\end{figure}
When the SNR satisfies the basic detection condition, i.e., $\text{SNR}\ge 1$, the UHTD can provide a constraint on the axion coupling constant:
\begin{align}\label{4.7}
    g_{a\gamma\gamma}^{(1)}\ge\sqrt{\frac{\hbar\omega_a/\tau }{2\kappa t_{in}\Delta f}}
\end{align}
where $\kappa=P_{\text{rf}}^{(2)}/g_{a\gamma\gamma}^2, ~P_{\text{rf}}^{(2)}=g_{a\gamma\gamma}^2|a_0|^2\omega_aQ_l\bar{B}^2 C_lV/2\mu_0$.

For conventional HTDs, which directly detect $P_{\text{rf}}^{(2)}$, the SNR is mainly affected by shot noise and thermal noise and can be written as
\begin{align}
    {\rm SNR}=\frac{P_{\text{rf}}^{(2)}\sqrt{t_{in}\Delta f}}{P_{\rm shot2}+P_{T}},
\end{align}
where $P_{\text{shot2}}=(P_{\text{rf}}^{(2)}\hbar\omega_a/\tau )^{1/2}$, $\tau=10^{-4}$ is the detector response time. 
Aagin, with $\rm SNR \ge 1$, the constraint on the axion coupling constant 
$g_{a\gamma\gamma}$
reads: 
\begin{align}\label{4.9}
g_{a\gamma\gamma}^{(2)}\ge \frac{1}{2\kappa\sqrt{t_{in}\Delta f}}\left(\sqrt{\frac{\kappa\hbar\omega_a}{\tau}}+\sqrt{\frac{\kappa\hbar\omega_a}{\tau}+4\kappa P_{T}}\right).
 \end{align}
 With the typical parameters~\cite{ADMX}: $\rho_a=0.5\times10^{-24}~\text{g/cm}^3$, $\bar{B}=8~\text{T}$, $C_l=1$, $\Delta f=100~\text{Hz}$, $t_{\text{in}}=100~\text{s}$, $\tau=10^{-4}~\text{s}$. The quality factor $Q$ of a resonant cavity is determined by both the intrinsic quality factor $Q_i$ and the coupling quality factor $Q_c$, satisfying the relation $Q = Q_i Q_c / (Q_i + Q_c)$. Since temperature variations significantly affect the intrinsic losses of the cavity, we assume the following values for $Q_i$: at $T = 300\,\mathrm{K}$, $Q_i = 10^5$; at $T = 4\,\mathrm{K}$, $Q_i = 3\times10^5$. The coupling quality factor is fixed at $Q_c = 10^5$ and is taken to be independent of temperature. Combining the UHTD sensitivity (Eq.~\eqref{4.7}), the conventional HTD sensitivity (Eq.~\eqref{4.9}), the KSVZ~\cite{A2023,SS2022} and DFSZ~\cite{PRB1981,SJNP1980} predictions, and the current experimental constraints, we present the axion sensitivity curves in Fig.~\ref{fig3}. Overall, the UHTD demonstrates superior sensitivity compared to the conventional scheme. In the typical GHz band, numerical calculations yield a three‑fold improvement (0.3 orders of magnitude) over the conventional method. In the MHz regime, where the conventional detector is dominated by the standard quantum limit (SQL), the figure shows that the sensitivity gain reaches nearly one order of magnitude.

In addition, the detectable signal amplitude is enhanced. In contrast, the gain provided by the UHTD is:
\begin{align}
    G=\frac{P_{\text{rf}}^{(1)}}{P_{\text{rf}}^{(2)}}=\sqrt{\frac{2P_{\text{rf}}^{(0)}}{P_{\text{rf}}^{(2)}}},
\end{align}
since $P_{\text{rf}}^{(0)} \gg P_{\text{rf}}^{(2)}$, the signal amplitude gain provided by this upgraded axion detection scheme is substantial.  

\begin{figure}
    \centering
    \includegraphics[width=0.65\linewidth]{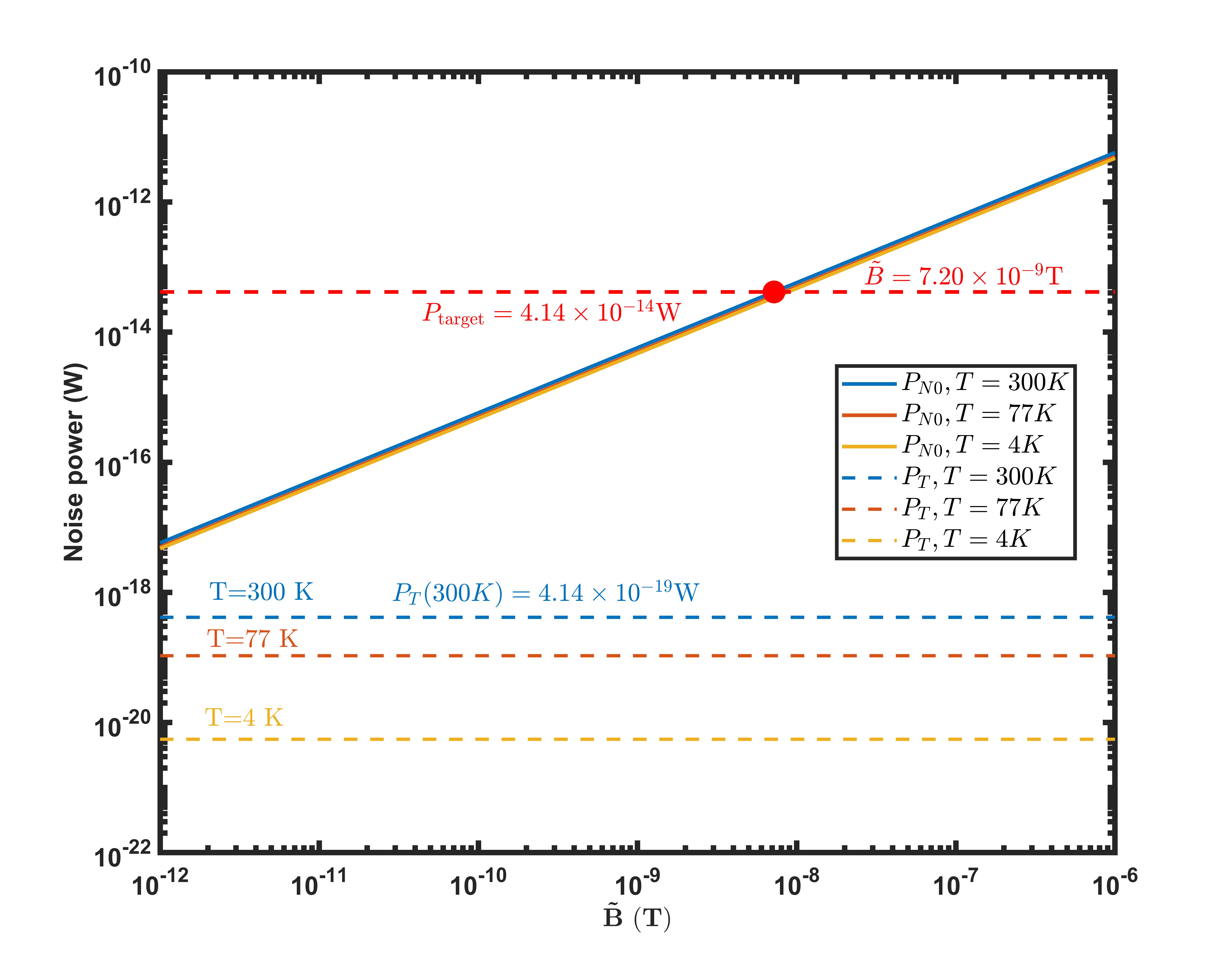}
    \caption{Shot‑to‑thermal noise power balance as a function of $\tilde{B}$.
}
    \label{fig4}
\end{figure}
Most importantly, in UHTD, other noises are strongly suppressed. Thermal noise, dark current, background noise, axion second-order shot noise, etc., are far smaller than the shot noise induced by the local oscillator signal at $\tilde{B}=10^{-6}\,\text{T}$. The quality factor of the resonant cavity is affected by temperature. The corresponding quality factors at temperatures of 300 K, 77 K, and 4 K are $5\times10^4$, $6.67\times10^4$, and $7.5\times10^4$, respectively.
Fig.\ref{fig4} illustrates the comparative relationship between the shot noise $P_{\text{N0}}$ and the thermal noise $P_T$ at various temperatures as a function of the magnetic field $\tilde{B}$. The colored solid lineS represents the shot noise ($P_{\text{N0}}$) induced by $\tilde{B}$, whereas the colored dashed lines denote the thermal noise ($P_T$) at 300K, 77K and 4K respectively. The red dashed line represents the thermal noise suppression threshold, and its power value is $P_{\text{target}}=5P_T(300\rm K)$. When the power of shot noise $P_{\text{N0}}$ exceeds this value, it's considered that the effect of thermal noise can be ignored. In Fig.~\ref{fig4}, we can see that the excited magnetic field value corresponding to this threshold power is $\tilde{B}=7.2\times10^{-9}\text{T}$.
At $\tilde{B}=10^{-6}\text{T}$, the local-oscillator-induced shot noise, originating from the quantum fluctuations of the magnetic field, overwhelmingly surpasses the thermal noise. This transition establishes the quantum-noise-limited regime, demonstrating that the detection system has reached the standard quantum limit (SQL). Consequently, the system performance becomes intrinsically temperature-insensitive.

 }

\section{Conclusions and Discussions.}
Given the existing HTD well-developed has not yet realized the desired EMR detection of the axions
in RF- and microwave bands, here we propose a UHTD to enhance the detection sensitivity. Specifically, we show that the detection of the existing HTD, which is biased only by a high SMF, is just the 2nd-order axion-photon energy signal; however, by additionally applying a transverse RF-excited magnetic field, the 1st EMR signal of the axions could be generated due to the 1st-order axion-photon interference energy signal. {\color{blue}Therefore, we argue that when the amplitude $\tilde{B}$ of the applied RF-excited magnetic field is set to $1\ \mu\text{T}$, the detection sensitivity of the proposed UHTD can be improved by 0.3 to 1 order of magnitude compared to that of conventional HTDs. Furthermore, the UHTD enhances the detectable axion signal amplitude, leading to a detectable power gain of approximately $G \sim \sqrt{2P_{\text{rf}}^{(0)} / P_{\text{rf}}^{(2)}}$. At $\tilde{B}=10^{-6}\text{T}$, the local‑oscillator‑induced shot noise, originating from quantum fluctuations of the magnetic field, far exceeds the thermal noise. This transition establishes a quantum‑noise‑limited regime, indicating that the detection sensitivity of  system has reached its quantum limit. Consequently, the system performance becomes intrinsically temperature‑insensitive, except its quality factor is still dependent of system.}

We now discuss the feasibility of the proposed UHTD. First, we argued that various robust filtering techniques could be utilized to implement the signal detection, although the strength of the generated 1st EMR signal is significantly weaker than that of the zeroth-order applied RF- or microwave signal (e.g., $P^{(0)}_{\rm rf}=1.19\times10^{-2}\rm W$ for $\tilde{B}=1\rm ~\mu T$ ). However, the latter one can be served, in principle, as an additional noise with the known frequency $\omega_B$, while the frequency of the 1st EMRs of axion could be physically generated as $\omega_a\pm\omega_B$. 
Next, by increasing the cooling power and using the materials with sufficiently good thermal conductivity, the unwanted heating effect of the strong zeroth-order EM background noise could be avoided effectively. Additionally, although the analysis demonstrated here is just based on an idealized one-dimensional configuration, the proposal could be directly applied to the practical three-dimensional cylindrical cavity detectors.

Anyway, given the transverse RF-excited magnetic field technique has been widely applied in the usual magnetic resonance imaging field, it is particularly desirable to upgrade the existing HTD to significantly enhance the detection sensitivity of the axion field by additionally applying the transverse RF-excited magnetic field for axion field detection is particularly desirable. 

\begin{acknowledgments}
This work was partially supported by the National Natural Science Foundation of China (Grant No. P110325G02011), the National Natural Science Foundation of China (Grant No. 12505083), the Natural Science Foundation of Sichuan Province (Grant No. 2025ZNSFSC0857), and the Natural Science Foundation of Sichuan Province (Grant No. 2025ZNSFSC0858).
\end{acknowledgments}





\end{document}


\maketitle
\flushbottom

\begin{appendix}

 \section{The signals generated in the conventional HTDs for axion fields detection}
 
In this section we first review briefly how the axion fields can be detected with the existing HTDs and then discuss the limitation in the previous approaches by additionally applying the rf signals to ex-situ coherently amplify the original EMR signals for improving their sensitivity. 
It is well-known that, in the usual HTD, which is biased by a high stationary magnetic field ${\bm B}^{(0)}$, the axion-modified Maxwell equation~\cite{NPB2024}:
\begin{equation}\label{eq2}
\left\{
\begin{aligned}		   
&\nabla\cdot\bm{E}(t)=\frac{1}{\varepsilon_0}g_{a\gamma\gamma}\bm{B}^{(0)}\cdot\nabla a,\\[3pt]
&\nabla\times\bm{B}(t)-\frac{1}{c^2}\partial_t\bm{E}(t)=-\mu_0g_{a\gamma\gamma}\bm{B}^{(0)}\partial_ta,\\
\end{aligned}
\right.
\end{equation}
can be obtained from the Lagrangian density. Here, $a=a({\bm{x}},t)$ is the axion field, $ \varepsilon_0$ is permittivity and $\mu_0$ the permeability in vacuum; ${\bm B}(t)$ and ${\bm E}(t)$ are the magnetic- and electric field densities of the EMR signal generated in the HTD, respectively.
Without the axion field and neglecting the background EM noise, the  electromagnetic field in the HTD cavity can be expressed as $E(t=0)=E^{(0)}=0$ and $B(t=0)=B^{(0)}$. Now, if an axion field passes through the HTD's cavity, an EMR signal with $E(t>0)\neq 0$ and $B(t>0)>B^{(0)}$ can be generated for the detection.

Under the axion field driving, the axion-modified Maxwell Eq.~\eqref{eq2} can be rewritten as
\begin{equation}\label{eq4}
\left\{    
\begin{aligned}		
&(-\nabla^2+\dfrac{1}{c^2}\partial^2_t)\phi            
=\dfrac{1}{\varepsilon_0}\rho_{\rm eff},        
\\		
&(-\nabla^2+\dfrac{1}{c^2}\partial^2_t)\bm{A}            
=\mu_0\bm{j}_{\rm eff},    
\end{aligned}
\right.
\end{equation}
where $\rho_{\rm eff}=g_{a\gamma\gamma}\bm{B}^{(0)}\cdot\nabla a$ is the axion-induced effective charge density, and $\bm{j}_{\rm eff}=g_{a\gamma\gamma}(\bm{E}^{(0)}\times\nabla a-\bm{B}^{(0)}\partial_ta)$ is the effective current density. Above, $\phi$ and ${\bm{ A}}$ are the electromagnetic potentials, which are satisfied by the Lorenz gauge condition. 
With the cavity's discrete spatial mode functions: $\bm{u}_{l}(x), l=1,2,...$, which satisfy the following orthogonality, normalization, and completeness relations:
\begin{equation}\label{eq10}
    \left\{
\begin{aligned}
		&\nabla\cdot(\varepsilon_0 \bm{u}_l)=0,\nabla\times(\frac{1}{\mu_0}\nabla\times\bm{u}_l)-\varepsilon_0 \omega_l^2\bm{u}_l=0, \\
		&\bm{n}\times\bm{u}_l|_s=0,\int_Vd^3x\varepsilon_0\bm{u}_l(\bm{x})\cdot\bm{u}_{l'}(\bm{x})=\delta_{ll'},
\end{aligned}
\right.
\end{equation}
the electromagnetic vector
potential ${\bm A}(\bm{x}, t)$ can be expanded as~\cite{SA2022,RSI2021}: ${\bm A}(\bm{x}, t)=\sum_l {\bm u}_l(\bm{x})\psi_l(t)$,
where $\psi_l(t)$ is the time-dependent function. For the case without any source driving, it satisfies the following dynamical equation:
\begin{align}\label{eq11}
	(\frac{d^2}{dt^2}+\Gamma_l\frac{d}{dt}+\omega_l^2)\psi_l(t)=0,
\end{align}
where $\Gamma_l=\omega_l/Q_l$
is the energy dissipation rate of the $l$th mode with $Q_l=\omega_l/\Gamma_l$ being its quality factor. 

Specifically, when an approximately isotropic axion wave, i.e.,  $\nabla a\ll\partial_ta$ and thus
$a(\bm{x},t)=Re(a_0e^{-i(\omega_at+\phi_a)})$ (with $\phi_a$
 being the random phase within the range 
$[0,2\pi)$~\cite{2018axionphase,QiaoliYang}) and $\rho_{\rm eff}\sim 0$, passes through a cavity biased by a
high SMF  
$\bm {B}^{(0)}=\bar{B}(\bm{x})$ (which is assumed to be along the $y$-direction),
then an effective current:
\begin{align}\label{eq12}
	\bm{j}_{\rm eff}(\bm{x},t)=-g_{a\gamma\gamma}\bar{B}(\bm{x})\partial_ta,
\end{align}
would be generated inside the cavity. Its spatial part can be written as: $\bm{j}_{\rm eff}(\bm{x})=\sum_lb_l\bm{u}_l(\bm{x})$, where
\begin{align}
b_l=\int_V\bm{j}_{\rm eff}(\bm{x})\cdot\bm{u}_l^*(\bm{x})d^3x=-g_{a\gamma\gamma}\int_V\bar{B}(\bm{x})\cdot\bm{u}_l^*(\bm{x})d^3x.
\end{align}
If $b_l\gg b_m\sim 0$, which refers to that the $l$-th mode in the cavity is excited by the effective current, then the time-dependence of the $l$-th mode function should be modified as:
\begin{align}\label{eq13}
		(\frac{d^2}{dt^2}+\Gamma_l\frac{d}{dt}+\omega_l^2)\psi_l&= {\rm Re}(-i\omega_aa_0e^{-i(\omega_at+\phi_a)}),
\end{align}
from Eq.~\eqref{eq11}. 
The solution of Eq.~\eqref{eq13} can be expressed specifically as:
\begin{align}\label{Eq.3.0}
\psi_l(t)=\exp\left[(-\Gamma_l/2\pm i\sqrt{\omega_l^2-\Gamma^2/4})t\right]+{\rm Re}\left(\frac{-ia_0\omega_ae^{-i(\omega_at+\phi_a)}}{(\omega_l^2-\omega_a^2-i\omega_l\Gamma_l)}\right),
\end{align}
where the first item can be omitted, as it will decay to zero for $t\rightarrow\infty$. As a consequence, 
we have
\begin{align}
&\bm{A}_l(\bm{x},t)=b_l\bm{u}_l(\bm{x})\psi_l(t)\nonumber\\
&=a_0g_{a\gamma\gamma}\bm{u}_l(\bm{x})\left(\int_V\bar{B}(\bm{x})\cdot\bm{u}_l(\bm{x})d^3x\right) {\rm Re}\left(\frac{i\omega_ae^{-i(\omega_at+\phi_a)}}{(\omega_l^2-\omega_a^2-i\omega_l\Gamma_l)}\right),
\end{align}
and thus the  electric- and magnetic field intensities of the response signals of the passing axion field can be calculated as $\bm{E}(\bm{x},t)=-\partial_t\sum_l \bm{A}_l(\bm{x},t)$,
 and $   \bm{B}=\nabla\times\sum_l\bm{A}_l(\bm{x},t)$, respectively. Up to the first-order approximation, they read 
\begin{align}\label{11}
    |\bar{E}^{(1)}_y|&=1.0095\times10^{-10}\left(\frac{V}{m}\right)\left(\frac{g_\gamma}{0.36}\right)\left(\frac{6\times10^{15}(eV)^2}{f_am_a}\right)\nonumber\\
&\times\left(\frac{\rho_a}{\frac{1}{2}\times10^{-24}g/cm^3}\right)\left(\frac{\bar{B}}{8T}\right)\left(\frac{Q_l}{10^4}\right),
\end{align}
and
\begin{align}\label{35}
    |\bar{B}^{(1)}_x|&=3.365\times10^{-17}T\left(\frac{g_\gamma}{0.36}\right)\left(\frac{6\times10^{15}(eV)^2}{f_am_a}\right)\nonumber\\
&\times\left(\frac{\rho_a}{\frac{1}{2}\times10^{-24}g/cm^3}\right)\left(\frac{\bar{B}}{8T}\right)\left(\frac{Q_l}{10^4}\right),
\end{align}
respectively. Above, we assume that the response is resonant, i.e., $\omega_a=\omega_l$, and the  experimental parameters are set typically as~\cite{SJNP1980}:
$g_\gamma=0.36$, $\rho_a=2^{-1}\times10^{-24}$ $\rm g/cm^3$,
	$\bar{B} =8$ T, $ V=1$~$\rm m^3$, $Q_l=10^4$, $\omega_a=2\pi\times 1$~GHz, and $f_am_a=6\times10^{15}\rm ~(eV)^2$. 
    Obviously, both the electric- and magnetic field intensities of the response signals of the passing axion field, generated in the usual HTD, are virtually undetectable; even with the most sensitive electric
~\cite{NC2024,RPP2023}
    and magnetic field detectors~\cite{ARBE2007} available today, they are  still a few orders of magnitude away from the sensitivities required to effectively probe these significantly weak signals. Therefore, the energy detection of the electromagnetic response signal of the passing axion field is necessary. 

    To this end, let us estimate below the signal power for the resonant response. Physically, the energy flux $\bm{S}^C_{a\rightarrow\gamma}=\bm{E}_l\times\bm{B}_l/\mu_0$ of the responded $l$-th mode electromagnetic field in the cavity can be calculated accordingly. 
As a result,  
the detectable power of the axion response signal, for the resonant response with $\omega_l=\omega_a$, can be expressed as: 
\begin{align}\label{eq2.0}
    P^{(2)}_{a\rightarrow\gamma}
=g_{a\gamma\gamma}^2\times\frac{2\rho_a\omega_a }{m_a^2\mu_0}\bar{B}^2Q_lC_lV,
\end{align}
with $C_l=(\int_Vd^3x\bar{B}(\bm{x})\cdot\bm{e}_l(\bm{x}))^2/(\bar{B} ^2V)$ being the form factor of the $l$-th mode of the electromagnetic field.
The detectable power is obviously  proportional to the square of $g_{a\gamma\gamma}$.

However, all the experimental detections by using the existing HTDs still provide a zero result to date. Therefore, upgrading the usual HTD-type installations to further improve their achievable detection sensitivities for capturing the weaker electromagnetic response signals of the axion fields, corresponding to the smaller axion-photon coupling parameter $g_{a\gamma\gamma}$, is particularly desirable.

\section{Using heterodyne-variance-based
detection in the usual HTDs}

Given the detectable axion-photon coupling strength, i.e., the value of $g_{a\gamma\gamma}$ parameter, is very limited, one must either continue to increase the strength of the steady-state SMF for generating the stronger electromagnetic response signal (as Eq.~\eqref{eq2.0} shows that the power of such an EMRs is proportional to increase the strength of SMF $\bar{B}$), or coherently amplify the amplitude of the generated electromagnetic response signal for improving its detectability.
Given the former approach is very limited by the high manufacturing cost for SMF, Ref.~\cite{PRD2023} proposed an approach by actively injecting the external photons into the HTD to coherently amplify the responded signal generated in the usual HTD. 

Indeed, following Ref.~\cite{PRD2023}, such an amplification is practically ex-situ, as it is just used to drive the cavity mode, i.e., 
\begin{align}\label{Eq1.0}
    \frac{d^2\psi_l'}{dt^2}+\Gamma_l\frac{d\psi_l'}{dt}+\omega_l^2X=E_d \cos(\omega_d t),
\end{align}
where $E_d \cos(\omega_d t)$ is an applied RF-driving source, with amplitude $E_d$ and frequency $\omega_d$, $\psi_l$ is the amplitude of the driven mode of the cavity with $\omega_l$ and $\Gamma_l=\omega_l/{Q_l}$ being its frequency and the energy dissipation coefficient, respectively. 
The solution of this equation Eq.~\eqref{Eq1.0} can be specifically written as:
\begin{align}
    \psi_l'(t)&=A_1e^{-\Gamma_l t/2}\cos (\sqrt{(\omega_l^2-\Gamma^2/4)}t+\phi_1)\nonumber\\
    &+E_d\frac{\omega\Gamma\sin (\omega_d t)+(\omega_l^2-\omega_d^2)\cos (\omega_d t)}{(\omega_l^2-\omega_d^2)^2+\omega_d^2\Gamma^2},
\end{align}
wherein the first term can be neglected within the sufficiently long driven duration with $\Gamma t\gg 1$. As a consequence, the total amplitude of the excited $l$-th cavity mode can be superposed as:
\begin{align}
X_l(t)&=\psi_l(t)+\psi'_l(t)\nonumber\\&=E_d\frac{\omega_d\Gamma\sin (\omega_d t)+(\omega_l^2-\omega_d^2)\cos (\omega_d t)}{(\omega_l^2-\omega_d^2)^2
+\omega_d^2\Gamma^2}\nonumber\\
&+E_a\frac{\omega_a\Gamma\sin (\omega_a t)+(\omega_l^2-\omega_a^2)\cos (\omega_a t)}{(\omega_l^2-\omega_a^2)^2+\omega_a^2\Gamma^2},
\end{align}
where $E_a$ is the amplitude of first order axion electromagnetic response, $\phi_d$ is the phase of the f-driving source. Consequently, the time-averaged power of 
the detectable $l$-th mode signal can be enhanced as:
\begin{align}
    \langle P_T\rangle&=\langle X_l(t)\frac{d X_l(t)}{dt}\rangle\nonumber\\&\propto \alpha\overline{[E_d\sin (\omega_dt+\phi_d)+E_a\cos (\omega_at+\phi_a)]^2}\nonumber\\
    &=\frac{1}{2} (E_d^2+E_a^2)+2(E_d\sin (\omega_dt+\phi_d))(E_a\cos (\omega_at+\phi_a))
\end{align}
for the resonant and coherent drivings with $\omega_a=\omega_d=\omega_l$, in the Refs.~\cite{PRD2023},  they assumed the phase difference of the axion and RF-driving source at $\pi/2$.  

The power of axion to
microwave photon conversion rate reads:
\begin{align}
    P_a=\frac{g_{a\gamma\gamma}^2\rho_a}{m_a}\bar{B}^2VC_l\frac{Q_cQ_a}{Q_c+Q_a}
\end{align}
where $Q_a$ is the quality factors of axion. The detectable power is obviously proportional to the square of
$g_{a\gamma\gamma}$.
The oscillating term is equal to zero if one integrates for a
longer time than the axion coherence time, estimated ~\cite{24}
to be on the order of 1 ms for 1 GHz axions. However, the
characteristic effect is to greatly enhance the variance of the
power of the particular frequency bin when we sample it
within a time shorter than the expected axion coherence
time.

The average number of transmitted photons on a power
meter per sampling $N_T$ is the sum of the signal photons
$N_s=P_a/(\hbar\omega_a\Delta f)$, the external photons $N_p$, and the noise photons $Nn =
N_{th}+ N_D$, where $N_{th}$ corresponds to the thermal noise
photons and $N_D$ is the dark count photons where all of
the sources follow Poisson statistics. The variance of the
transmitted photons for each sample, for a signal with
random phase, is
\begin{align}
    \hat{\sigma}^2(N_T) = N_{th} + N_D + N_s + N_p (1 + N_s + N_{th}).
\end{align}
We can observe that after introducing the photon beam, when using heterodyne-variance-based to detect the signal, the variance remains proportional to the square of the coupling constant $g_{a\gamma\gamma}$.

Additionally, in the test experiment, the signal
from a network analyzer was used as the probe, and a
random phase axion signal was generated from the signal
generator. The signal and the probe were synthesized with
a directional coupler just before injection into the cavity.
The network analyzer scanned the vicinity of the resonance
frequency of the test cavity, and the transmission was
repeatedly measured with an axion signal at the resonance
frequency each time. 
\begin{figure}
    \centering
    \includegraphics[width=0.9\linewidth]{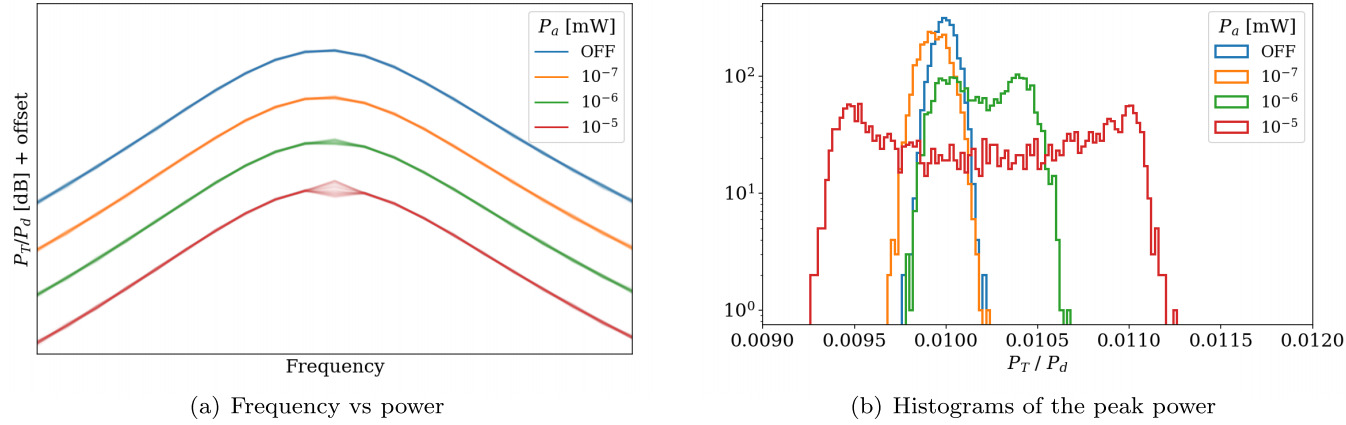}
    \caption{The test experiment results by using heterodyne-variance-based detection in the usual HTDs~\cite{PRD2023}.}
    \label{Afig1}
\end{figure}
The results are illustrated showing, in Fig.~\ref{Afig1}(a), multiple
traces of power spectra with increasing axion power at a
constant probe power and, in Fig.~\ref{Afig1}(b), the observed power
distribution at the peaks of the curves (axion resonance
frequency). A clear variance was observed at the resonance
frequency that when injected in, and the histogram shows that
the peaks appear at both ends for sufficient signal power,
indicating that it follows the distribution of a sine function~\cite{PRD2023}.

Specifically, one can see from  Fig.~\ref{Afig1} that, when the injected axion power signal reaches at least $10^{-6}$ mW, subtle changes can be observed in the variance plot and histogram. However, for the extremely weak axion-photon coupling, generating such a strong power of $10^{-6}$ mW is extremely challenging. 

\section{The EM response of the axion field passing through the proposed UHTD}

In the UHTD, besides the 
usual high SMF $\bar{B}$ applied along the $y$-direction, a RF-excited field is applied here to excite the cavity's magnetic resonant mode along the $x$ - direction with the amplitude being $\tilde{B}\ll \bar{B}$.  As a consequence, the background magnetic field in the existing HTD can be upgraded as:
\begin{align}\label{eq16}
	\bm{B}^{(0)}(\bm{x},t)=\bar{\bm{B} }\bm{e}_y+\tilde{B}\mathrm{Re}\left(e^{i(\bm{k}_B\cdot\bm{x}-\omega_{B}t)}\right)\bm{e}_x,
\end{align}
where $\tilde{B}$, $\bm{k}_B$ and $\omega_B$ is the amplitude, wave vector and frequency of the RF-excited field, respectively. Taking the background magnetic field Eq.~\eqref{eq16} into the axion-modified Maxwell equations shown in Eq.~\eqref{eq2}, the equations solution can be formally expressed as:
\begin{equation}\label{eq17}
    \left\{
\begin{aligned}
	\bm{B}=&\bm{B}^{(0)}+\bm{B}^{(1)}+\mathcal{O}^{(2)}(g_{a\gamma\gamma}),\\[3pt]
	\bm{E}=&\bm{E}^{(0)}+\bm{E}^{(1)}+\mathcal{O}^{(2)}(g_{a\gamma\gamma}),
\end{aligned}
\right.
\end{equation}
the superscripts: $(0)$ and $(1)$, indicate the effects related to the zeroth- and first-order (1st) responses of the axion-photon coupling, and the effects related to the second-order (2nd) and higher ones, i.e., $\mathcal{O}^{(2)}$, are safely neglected. 
Given the zeroth-order effect can be easily calculated by solving the conventional Maxwell equation in the absence of axion-photon coupling. The 1st EMR of the axion can be described by
\begin{equation}\label{eq20}
\renewcommand{\arraystretch}{1.5} 
    \left\{
\begin{aligned}	&\nabla\cdot\bm{E}^{(1)}=\frac{1}{\varepsilon_0}g_{a\gamma\gamma}\bm{B}^{(0)}\cdot\nabla a\sim 0,\\[3pt]
		&\nabla\bm{B}^{(1)}-\frac{1}{c^2}\partial_t\bm{E}^{(1)}=\mu_0\bm{j}^{(1)}_{\rm eff}(\bm{x},t),
\end{aligned}
\right.
\end{equation}
for the approximately isotropic axion wave, i.e., the axion field can be described as $a(\bm{x},t)\approx Re(a_0e^{-i(\omega_at+\phi_a)})$. 

Under the Lorenz gauge condition; $\bm{B}^{(1)}=\nabla\times\bm{A}^{(1)},\bm{E}^{(1)}=-\partial_t\bm{A}^{(1)}-\nabla \phi^{(1)},\nabla\cdot \bm{A}^{(1)}+\partial_t\phi^{(1)}=0$, we can conclude that:
$   -\nabla^2\bm{A}^{(1)}+\partial_t^2\bm{A}^{(1)}/c^2=\bm{j}_{eff}^{(1)},
$
When a RF-excited magnetic field is added perpendicular to the direction of the steady magnetic field, the resulting axion response effective current will become
\begin{align}
    \bm{j}_{\mathrm{eff}}^{(1)}=-\frac{1}{4c}g_{a\gamma\gamma}[\bar{B}\bm{e}_{y}+\tilde{B}\operatorname{Re}(e^{i(\bm{k}_{B}\cdot\bm{x}-\omega_{B}t)})\bm{e}_{x}]\times\partial_{t} a.
\end{align}

In the cavity, the electromagnetic vector potential $\bm{A}^{(1)}$ can be expanded in terms of a series of discrete modes: $\bm{A}^{(1)}=\sum_l\bm{u}_l(\bm{x})\psi_l(t)$. For the $l$-th cavity mode with the dissipation $\Gamma_l$, its response to the axion field, i.e., the signal transported along the $x$-direction can be divided into space part and time-dependent part. And the space part satisfies the orthogonality, normalization, and completeness relations. Then, the time-dependent part will satisfy the following equation:
\begin{align}\label{eq22}
   & (\frac{d^2}{dt^2}+\Gamma_l\frac{d}{dt}+\omega_l^2) 
\psi^{(1)}_{x,l}(t)
    =a_{lx}\Omega\mathrm{Re}(e^{-i(\omega_at+\phi_a)})\mathrm{Re}(e^{-i\omega_Bt}), 
\end{align}
with $\Omega=
cg_{a\gamma\gamma}\omega_a|a_0|/4$ and $a_{lx}=\int_V\tilde{B}(\bm{x})\cdot\bm{u}_l^*(\bm{x})d^3x $ is the overlap coefficient of the effective flow to cavity mode.  The electromagnetic vector potential $x$-component solution reads 
$\bm{A}^{(1)}_{x,l}(\bm{x},t)=\left(A^{(1)}_{x,l^+}(\bm{x},t)+A^{(1)}_{x,l^-}(\bm{x},t)\right)\bm{e}_x$, where
$\bm{A}^{(1)}_{x,l^\pm}(\bm{x},t)=\mathrm{Re}({ia_{lx}\Omega \bm{u}_l(\bm{x})e^{-i(\Delta_{aB}^\pm t+\phi_a)}}/\\(\omega_{l}^2-(\Delta_{aB}^\pm)^2-i\Gamma_{l}(\Delta_{aB}^\pm))$, 
and $\Delta^\pm_{aB}=\omega_a\pm\omega_B$.
Similarly, for the response along the $y$-direction, we have
\begin{align}\label{eq25}
   & (\frac{d^2}{dt^2}+\Gamma_l\frac{d}{dt}+\omega_l^2) 
\psi^{(1)}_{y,l}(t)
    =a_{ly}\Omega\mathrm{Re}(e^{-i(\omega_at+\phi_a)}),
\end{align}
where, $a_{ly}=\int_V\bar{B}(\bm{x})\cdot\bm{u}_l^*(\bm{x})d^3x $, whose solution reads
$\bm{A}^{(1)}_{y,l}(\bm{x},t)=a_{ly}\Omega\bm{u}_{l}(\bm{x})
\mathrm{Re}(ie^{-i(\omega_at+\phi_a)}/\\(\omega_{l}^2-\omega_a^2-i\Gamma_{l} \omega_a))\bm{e}_y$. According to the relationship between electromagnetic field and the magnetic vector potential, the electromagnetic Eq.~\eqref{eq17} in the cavity can be specifically expressed as:
\begin{equation}\label{eq26}
    \left\{
\begin{aligned}
	\bm{E} =& \tilde{E}_x^{(1)}\bm{e}_x +\left[E_y^{(0)} +\bar{E}_y^{(1)}\right]\bm{e}_y,\\[3pt]
	\bm{B}=& \left[\tilde{B}_x^{(0)} 
                        +\bar{B}_x^{(1)} \right]\bm{e}_x  + \left[\bar{B} + \tilde{B}_y^{(1)}\right]\bm{e}_y,
\end{aligned}
    \right.
\end{equation}
with
\begin{equation}
    \left\{
\begin{aligned}\label{eq19}
	&\bm{E}_y^{(0)}(\bm{x},t)=-\mathrm{Re}(c\tilde{B}e^{i(\bm{k}_B\cdot\bm{x}-\omega_Bt)})\bm{e}_y,\\[3pt]
&
 \bm{B}^{(0)}(\bm{x},t)=\bar{\bm{B} }\bm{e_y}+\mathrm{Re}(\tilde{B}e^{i(\bm{k_B}\cdot\bm{x}-\omega_{B}t)})\bm{e}_x,
\end{aligned}
\right.
\end{equation}
for the zeroth-order EMR signal,
\begin{equation}\label{eq27}
    \left\{
\begin{aligned}
		\tilde{E}^{(1)}_x(\bm{x},t)&=
a_{lx}\Omega\bm{u}_l(\bm{x})(\Delta_{aB}^+ \Phi_{l+}+\Delta_{aB}^-\Phi_{l-}),\\[3pt]
		\bar{E}^{(1)}_y(\bm{x},t)&=\omega_aa_{ly}\Omega\bm{u}_l(\bm{x})\Phi_l,
\end{aligned}
 \right.
\end{equation}
and
\begin{equation}\label{eq28}
    \left\{
\begin{aligned}
	\tilde{B}^{(1)}_y(\bm{x},t) = &-\bm{k}_la_{lx}\Omega\bm{u}_l(\bm{x})[ 
   \Phi_{l+}
    +\Phi_{l-}],\\[3pt]
		\bar{B}^{(1)}_x(\bm{x},t)=&-\bm{k}_la_{ly}
    \Omega\bm{u}_l(\bm{x})\Phi_l,
\end{aligned}
   \right.
\end{equation}
for the 1st EMR signal. Above, $\bm{k}_l=\omega_l/c$, $\Phi_{l\pm}=\mathrm{Re}\left(e^{-i(\Delta_{aB}^\pm t+\phi_a)}/[\omega_{l}^2-(\Delta_{aB}^\pm)^2-i\Gamma_{l}(\Delta_{aB}^\pm)]\right)$, and $\Phi_{l}=\mathrm{Re}\left(e^{-i(\omega_at+\phi_a)}/[\omega_{l}^2-\omega_a^2-i\Gamma_{l}\omega_a]\right)
$.
Obviously, these intensities are proportional to the weak axion-photon coupling strength $g_{a\gamma\gamma}$. 

Under the full resonance condition: $\omega_{l}\approx \omega_a\approx\omega_B$, the axion's responded electric field, contributed by the RF-excited magnetic field   
$\tilde{B}=10^{-6}$ T, can be expressed as
\begin{align}\label{33}
    |\tilde{E}^{(1)}_x|=&1.5775\times10^{-16}\left(\frac{V}{m}\right)\left(\frac{g_\gamma}{0.36}\right)\left(\frac{6\times10^{15}(eV)^2}{f_am_a}\right)\nonumber\\
&\times\left(\frac{\rho_a}{\frac{1}{2}\times10^{-24}g/cm^3}\right)
    \left(\frac{\tilde{B}}{1~\mu T}\right)\left(\frac{Q_l}{10^4}\right),
\end{align}
and
\begin{align}\label{36}
    |\tilde{B}^{(1)}_y|=&5.258\times10^{-25}T\left(\frac{g_\gamma}{0.36}\right)\left(\frac{6\times10^{15}(eV)^2}{f_am_a}\right)\nonumber\\
&\times\left(\frac{\rho_a}{\frac{1}{2}\times10^{-24}g/cm^3}\right)\left(\frac{\tilde{B}}{1~\mu T}\right)\left(\frac{Q_l}{10^4}\right),
\end{align}
besides those (shown above in Eq.~\eqref{11} and ~\eqref{35}, respectively) contributed by the original high SMF. %
Obviously, in this scheme, the intensity of the additional response electromagnetic field generated by the interaction between the RF-excited magnetic field and the axion field is much weaker than that of the response electromagnetic field generated from the conventional HTD, and therefore the additional response electromagnetic field  cannot be detected still.
However, due to the RF-excited magnetic field driving, the 1st energy flux density of the axion response signal can be additionally generated as:
\begin{align}\label{eq29}
    \bm{S}^{(1)}_{rf}&=\frac{1}{\mu_0}\left(\tilde{E}_x^{(1)}\bar{B} -E_y^{(0)}\bar{B}_x^{(1)}-\bar{E}_y^{(1)}B_x^{(0)}\right)\bm{e}_z\nonumber\\
        &=-\frac{a_{lx}\Omega\bar{B}\bm{u}_l(\bm{x})}{\mu_0}[\Delta_{aB}^+ \Phi_{l+}+\Delta_{aB}^-\Phi_{l-}]\bm{e}_z\nonumber\\
&+\frac{\omega_l}{\mu_0}a_{ly}\Omega\bm{u}_l(\bm{x})\tilde{B}\Phi_l\mathrm{Re}\left(e^{i(\bm{k}_B\cdot\bm{x}-\omega_Bt)}\right)\bm{e}_z,
\end{align}
besides the zeroth-order energy flux, $\bm{S}^{(0)}_{rf}=E_y^{(0)}B_x^{(0)}\bm{e}_z/\mu_0$ (shown in Eq.~\eqref{eq19}), which is independent of axion. 
 
Continuously, the 2nd energy flux density of the present EMR signal of the axion field can be calculated as:
\begin{align}\label{37}
		\bm{S}^{(2)}_{rf}&=\frac{\Gamma_l}{\mu_0}\bm{E}^{(1)}\times\bm{B}^{(1)}=\frac{\Gamma_l}{\mu_0}\left(\tilde{E}_x^{(1)}\tilde{B}_y^{(1)}-\bar{E}_y^{(1)}\bar{B}_x^{(1)}\right)
		\nonumber\\
        &=\frac{\Gamma_l}{\mu_0}\left(\partial_t A^{(1)}_{x,l}\partial_z A^{(1)}_{x,l}+\partial_tA^{(1)}_{y,l}\partial_zA^{(1)}_{y,l}\right)\nonumber\\
		=&-\frac{\Gamma_l}{\mu_0}\bm{k}_la_{lx}^2\Omega^2\bm{u}_l^2(\bm{x})\left[(\Delta_{aB}^+\Phi_++\Delta_{aB}^-\Phi_{l-})(\Phi_++\Phi_{l-})\right]\nonumber\\
        &+\frac{\Gamma_l}{\mu_0}\bm{k}_l\omega_aa_{ly}^2\Omega^2\bm{u}_l^2(\bm{x})\Phi_l^2,
\end{align}
and its time-average reads
\begin{align}\label{36}
\langle\bm{S}^{(2)}_{rf}\rangle=cg_{a\gamma\gamma}^2|a_0|^2\omega_aQ_l(\bar{B}^2+\tilde{B}^2) C_lV\bm{e}_z,
\end{align}
which is obviously  proportional to the square of axion-photon coupling strength (as $\Omega=
cg_{a\gamma\gamma}\omega_a|a_0|/4$ defined in Eq.~\eqref{eq22}), the
squares of amplitudes of the biased high SMF (as $a_{ly}\sim\bar{B} $ shown in Eq.~\eqref{eq25}), and the additionally applied RF-excited field (as $a_{lx}\sim\tilde{B}$ shown in Eq.~\eqref{eq22}. Since $\bar{B}\gg\tilde{B}$, the 2nd energy response is mainly contributed by the biased high SMF. In contrast, the 1st energy response of the passing axion field is only generated by the applied RF-excited field, i.e., for $\tilde{B}=0$ and thus $a_{lx}=0$, we have  $S^{(1)}_{rf}=0$. As we show below that, the first-order axion-photon energy response significantly enhances the axion-photon energy transferring rate, compared with the usual 2nd-order axion-photon energy response.

Physically, the phase factor might be treated as an arbitrarily unknown random variable, and thus if it influences the axion-photon energy conversion should be discussed. 
Following Refs.~\cite{2018axionphase,QiaoliYang}, an isotropic axion particle could be expressed as:
\begin{align}
    a_i(v_i,t)=\frac{\sqrt{2\rho_{a}/N_a}}{m_a}\cos\left(\omega_it+\phi_a\right),
\end{align}
where $N_a$ is the particle number for the local DM
density $\rho_{a}$, $\omega_i=m_a(1+v_i/2)$ with $v_i$ being the speed of the $i$th-axion particle and $\phi_a$ the unknown phase. In order to extend the axion field to the local DM density, one may change the $N_a$-particle distribution into a subset $\Omega_j$,  which contain the $N_j$ particles with speeds
between $v_j$ and $v_j + \Delta v$, with $\Delta v$ being small enough that the difference between their speeds can be neglected. As a consequence,
the contribution from all particles in the subset $\Omega_j$ is given by
\begin{align}
    a_j(t)=\sum_{i\in\Omega_j}\frac{\sqrt{2\rho_{a}/N_a}}{m_a}\cos\left(\omega
_jt+\phi_a\right).
\end{align}
Given that the phase is random, e.g., $\phi_a\in [0,2\pi)$, we get
$$
\sum_{i\in\Omega_j} \cos\left[\omega_jt+\phi_a\right]
    ={\rm Re}\left[{\rm \exp}\left(i\omega
_jt\right)\left(\sum_{i\in \Omega_j}{\rm \exp}(i\phi_a)\right)\right],
$$
with $\sum_{i\in \Omega_j}{\rm exp}(i\phi_a)=\alpha_je^{i\phi_a}$.   As a consequence, the amplitudes $\alpha_j$ can be described by the Rayleigh distribution
~\cite{2018axionphase}, i.e.,
\begin{align}\label{C4}
    P[\alpha_j]=\frac{2\alpha_j}{N_a^j}e^{-\alpha_j^2/N_a^j}.
\end{align}
With such a assumption, the 1st energy flux density of the axion's response signal, shown in Eq.~\eqref{eq29} for $\phi_a=0$, should be replaced as the $\phi_a$-dependent one $S_{rf}^{(1)}(\phi_a)$.   
Consequently, under the resonant driving, i.e., 
$\omega_a=\omega_B=\omega_l$, the time-average of $S_{rf}^{(1)}$  should be expressed as \begin{align}\label{C6}
     \langle \bm{S}_{rf}^{(1)}(\phi_a)\rangle&=\lim_{T\rightarrow\infty}\frac{1}{T}\int_0^T\frac{\bm{e}_z}{\mu_0}\left(\tilde{E}_x^{(1)}\bar{B} -E_y^{(0)}\bar{B}_x^{(1)}-\bar{E}_y^{(1)}B_x^{(0)}\right)dt\nonumber\\
&=cg_{a\gamma\gamma}|a_0|\bar{B}\tilde{B}Q_l C_lV\cos{(\phi_a)}\bm{e}_z.
 \end{align}
 where  $C_l=(\int_Vd^3x\bar{B}(\bm{x})\cdot\bm{u}_l(\bm{x})\int_Vd^3x\tilde{B}(\bm{x})\cdot\bm{u}_l(\bm{x}))/(\bar{B} \tilde{B}V)$ is the form factor of the $l$-th mode electromagnetic field in the cavity.
By making the statistical average on the random phase with the distribution function ~\eqref{C4}, we have \begin{align}\label{C7}
     \overline{\langle \bm{S}_{rf}^{(1)}(\phi_a)\rangle} =cg_{a\gamma\gamma}|a_0|\bar{B}\tilde{B}Q_l C_lV\overline{\cos(\phi_a)}\bm{e}_z,
 \end{align}
 where $\overline{\cos(\phi_a)}$ is the statistical average of $\cos(\phi_a)$ for the unknown stochastic phase $\phi_a\sim [0,2\pi)$~\cite{2018axionphase,QiaoliYang}. 
 As a result, it cannot be directly detected, as $\overline{\cos(\phi_a)}$ is zero for the long-time averages.
 
 Alternatively, we show below that the stable output of 1st-order axion response signals can be implemented by using the usual IQ-mixer  modulation technique, although the phase of the 1st-order axion response signal is still random~\cite{IQ1,IQ2,IQ3}. in fact, the 1st energy flux density shown in Eq.~\eqref{eq29} can be rewritten as:
 \begin{align}\label{St}
    S^{(1)}_{rf}(t)=|S_{rf}^{(1)}|[\cos(\Delta_{aB}^+t+\phi_a)+\cos(\Delta_{aB}^-t+\phi_a)],
\end{align}
with $|S_{rf}^{(1)}|=ca_0g_{a\gamma\gamma}\omega_a^2\bar{B}\tilde{B}/(4\mu_0\Gamma\omega_a)$ being its amplitude.
Under the resonant condition, i.e., 
$\omega_a=\omega_B=\omega_l$, we have
$
S^{(1)}(t)\sim |S_{rf}^{(1)}|[\cos(2\omega_Bt+\phi_a)+\cos{(\phi_a)}]$, which can be split into two signals $S^{(1)}_1(t)$ and $S^{(1)}_2(t)$ with the same amplitude of $|S_{rf}^{(1)}|/2$. They are mixed with the local coherent electromagnetic signals  $L_1(t)\sim L\cos(2\omega_Bt)/2$ and $L_2(t)\sim L\sin(2\omega_Bt)/2$, where $L$ is the known amplitude. After the low-pass filtering, we get the $I(\phi_a)$-channel signal:
\begin{align}
    I(\phi_a)=\frac{1}{4}|S_{rf}^{(1)}|L\cos(\phi_a),
\end{align}
and the $Q(\phi_a)$-channel signal:
\begin{align}
    Q(\phi_a)=-\frac{1}{4}|S_{rf}^{(1)}|L\sin(\phi_a),
\end{align}
respectively. 
Finally, the amplitude of the 1st response signal, $|S_{rf}^{(1)}|$, can be given by the demodulation signal from the IQ mixer
\begin{align}\label{S1}
D= \sqrt{I^2(\phi_a)+Q^2(\phi_a)}=\frac{1}{4}|S_{rf}^{(1)}|L,
\end{align}
which is proportional to $g_{a\gamma\gamma}$ and independent of the signal phase.

\end{appendix}